\documentclass[twocolumn,preprintnumbers,amsmath,amssymb]{revtex4}

\usepackage{graphicx}
\usepackage{dcolumn}
\usepackage{bm}
\usepackage{rotating}

\usepackage{bbm}
\usepackage{verbatim}
\usepackage{afterpage}

\usepackage{color}
\definecolor{lightblue}{rgb}{0.2,0.2,0.7}
\definecolor{darkblue}{rgb}{0,0.25,0.5}
\definecolor{redbrown}{rgb}{0.875,0.25,0.125}
\definecolor{darkgreen}{rgb}{0,0.5,0}

\newcommand{\bra}[1]{\ensuremath{\langle #1 \vert}}
\newcommand{\ket}[1]{\ensuremath{\vert #1  \rangle}}

\renewcommand{\b}[1]{\ensuremath{\mathbf{#1}}}

\renewcommand{\l}{\ensuremath{\lambda}}

\newcommand{\KS}{\ensuremath{\text{KS}}}

\newcommand{\HF}{\ensuremath{\text{HF}}}

\begin{document}

\title{Rationale for a new class of double-hybrid approximations in density-functional theory}

\author{Julien Toulouse$^1$}\email{julien.toulouse@upmc.fr}
\author{Kamal Sharkas$^{1,2}$}
\affiliation{
$^1$Laboratoire de Chimie Th\'eorique, Universit\'e Pierre et Marie Curie and CNRS, 75005 Paris, France\\
$^2$Atomic Energy Commission of Syria, P.O. Box 6091, Damascus, Syria}
\author{Eric Br\'emond}
\author{Carlo Adamo}\email{carlo-adamo@chimie-paristech.fr}
\affiliation{
Laboratoire d'\'Electrochimie, Chimie des Interfaces et Mod\'elisation pour l'\'Energie, Chimie ParisTech and CNRS, 75005 Paris, France}

\date{\today}

\begin{abstract}
We provide a rationale for a new class of double-hybrid approximations introduced by Br\'emond and Adamo [J. Chem. Phys. {\bf 135}, 024106 (2011)] which combine an exchange-correlation density functional with Hartree-Fock exchange weighted by $\l$ and second-order M{\o}ller-Plesset (MP2) correlation weighted by $\l^3$. We show that this double-hybrid model can be understood in the context of the density-scaled double-hybrid model proposed by Sharkas {\it et al.} [J. Chem. Phys. {\bf 134}, 064113 (2011)], as approximating the density-scaled correlation functional $E_c[n_{1/\l}]$ by a linear function of $\l$, interpolating between MP2 at $\l=0$ and a density-functional approximation at $\l=1$. Numerical results obtained with the Perdew-Burke-Ernzerhof density functional confirms the relevance of this double-hybrid model.
\end{abstract}

\maketitle

The double-hybrid (DH) approximations introduced by Grimme~\cite{Gri-JCP-06}, after some related earlier work~\cite{ZhaLynTru-JPCA-04,ZhaLynTru-PCCP-05}, are increasingly popular for electronic-structure calculations within density-functional theory. They consist in mixing Hartree-Fock (HF) exchange with a semilocal exchange density functional and second-order M{\o}ller-Plesset (MP2) correlation with a semilocal correlation density functional:
\begin{eqnarray}
E_{xc}^{\text{DH}} &=& a_x E_{x}^{\HF} + (1-a_x) E_{x}[n]
\nonumber\\
&& + (1-a_c) E_{c}[n] + a_c E_{c}^{\text{MP2}},
\label{DH}
\end{eqnarray}
where the first three terms are calculated in a usual self-consistent hybrid Kohn-Sham (KS) calculation, and the last perturbative term is evaluated with the previously obtained orbitals and added \textit{a posteriori}. The two empirical parameters $a_x$ and $a_c$ can be determined by fitting to a thermochemistry database. For example, the B2-PLYP double-hybrid approximation~\cite{Gri-JCP-06} is obtained by choosing the Becke 88 (B) exchange functional~\cite{Bec-PRA-88} for $E_{x}[n]$ and the Lee-Yang-Parr (LYP) correlation functional~\cite{LeeYanPar-PRB-88} for $E_{c}[n]$, and the empirical parameters $a_x=0.53$ and $a_c=0.27$ are optimized for the G2/97 subset of heats of formation. Another approach has been proposed in which the perturbative contribution is evaluated with normal B3LYP orbitals rather than orbitals obtained with the weighted correlation density functional $(1-a_c) E_{c}[n]$~\cite{ZhaXuGod-PNAS-09,ZhaXu-IRPC-11}.

Recently, Sharkas, Toulouse, and Savin~\cite{ShaTouSav-JCP-11} have provided a rigorous reformulation of the double-hybrid approximations based on the adiabatic-connection formalism, leading to the {\it density-scaled one-parameter double-hybrid} (DS1DH) approximation
\begin{eqnarray}
E^{\text{DS1DH},\l}_{xc} &=& \l E_x^{\HF} + (1-\l) E_x[n] 
\nonumber\\
&&+ E_c[n] -\l^2 E_c[n_{1/\l}] + \l^2 E_c^{\text{MP2}},
\label{ExcDS1DH}
\end{eqnarray} 
where $E_c[n_{1/\l}]$ is the usual correlation energy functional evaluated at the scaled density $n_{1/\l}(\b{r})=(1/\l)^3 n(\b{r}/\l)$. This reformulation shows that only one independent empirical parameter $\l$ is needed instead of the two parameters $a_x$ and $a_c$. The connection with the original double-hybrid approximations can be made by neglecting the density scaling
\begin{eqnarray}
E_c[n_{1/\l}] \approx E_c[n],
\label{Ecnoscaled}
\end{eqnarray} 
which leads to the {\it one-parameter double-hybrid} (1DH) approximation~\cite{ShaTouSav-JCP-11}
\begin{eqnarray}
E^{\text{1DH},\l}_{xc} &=& \l E_x^{\HF} + (1-\l) E_x[n] 
\nonumber\\
&&+ (1-\l^2) E_c[n] + \l^2 E_c^{\text{MP2}}.
\label{Exc1DH}
\end{eqnarray} 
Equation~(\ref{Exc1DH}) exactly corresponds to the double-hybrid approximation of Eq.~(\ref{DH}) with parameters $a_x=\l$ and $a_c=\l^2$.

\begin{figure*}
\includegraphics[scale=0.3,angle=-90]{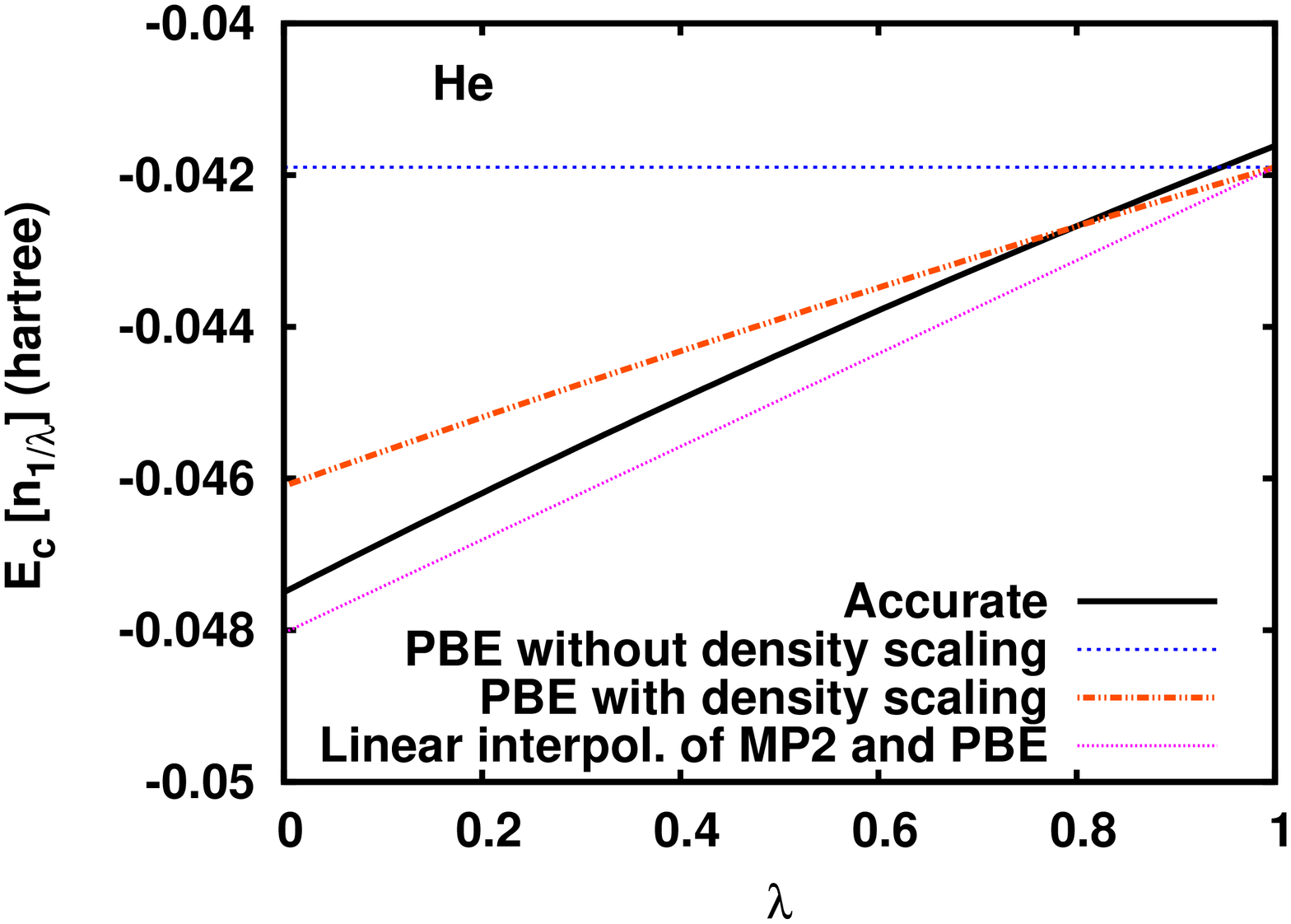}
\includegraphics[scale=0.3,angle=-90]{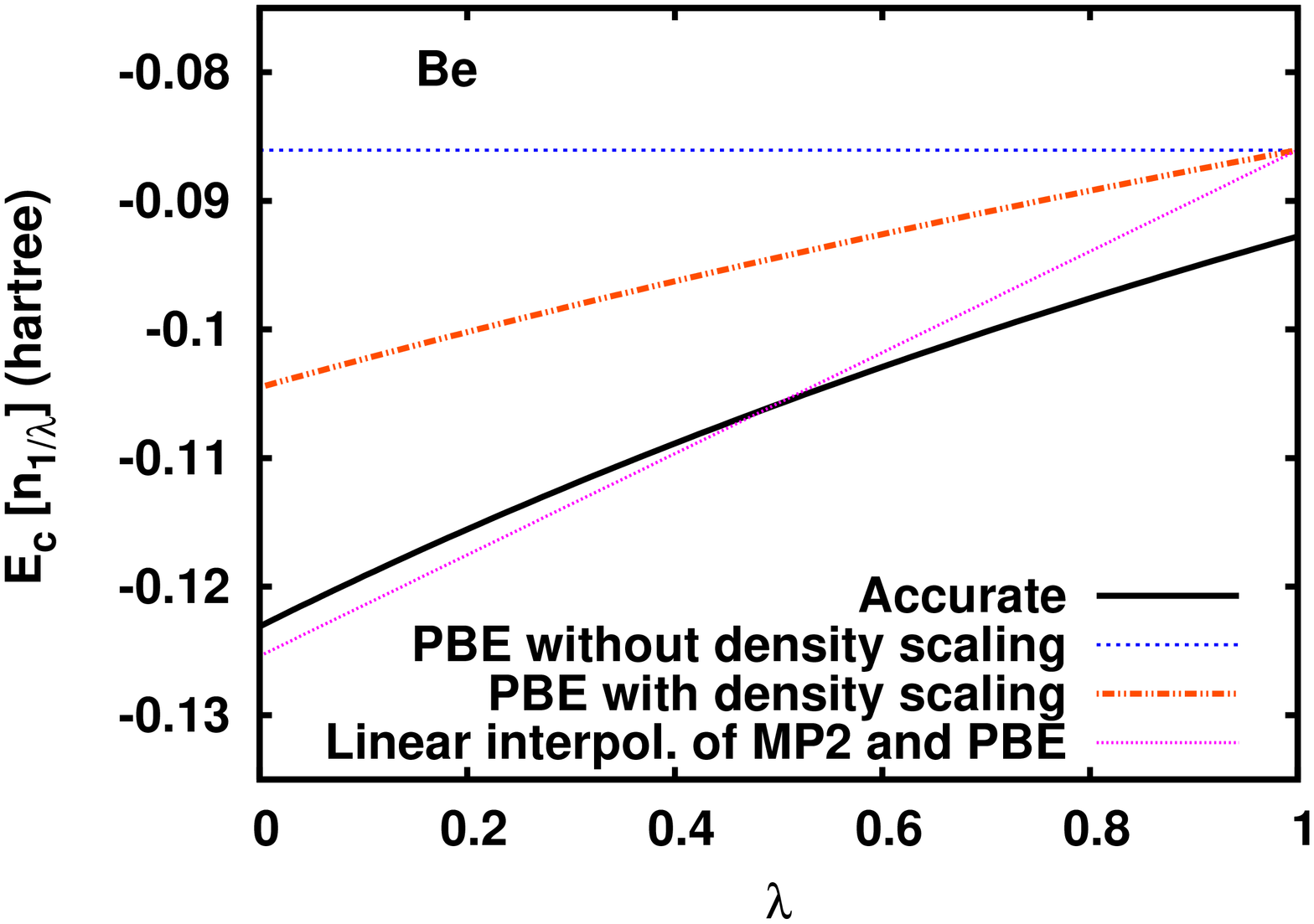}
\caption{(Color online) Density-scaled correlation energy $E_c[n_{1/\l}]$ for the He (left) and Be (right) atoms as a function of $\l$. Accurate calculations (from the parametrizations of Ref.~\onlinecite{ColSav-JCP-99}) are compared with different approximations: PBE without density scaling [Eq.~(\ref{Ecnoscaled})], PBE with density scaling (from the parametrizations of Ref.~\onlinecite{ColSav-JCP-99}), and linear interpolation between MP2 (with PBE orbitals) and PBE [Eq.~(\ref{Ecscaledlinear})]. The MP2 calculations for He and Be (including core excitations) have been performed with the cc-pV5Z and cc-pCV5Z basis sets~\cite{WooDun-JCP-94,PraWooPetDunWil-TCA-11}, respectively.
}
\label{fig:hebe}
\end{figure*}

\begin{figure*}
\includegraphics[scale=0.3,angle=-90]{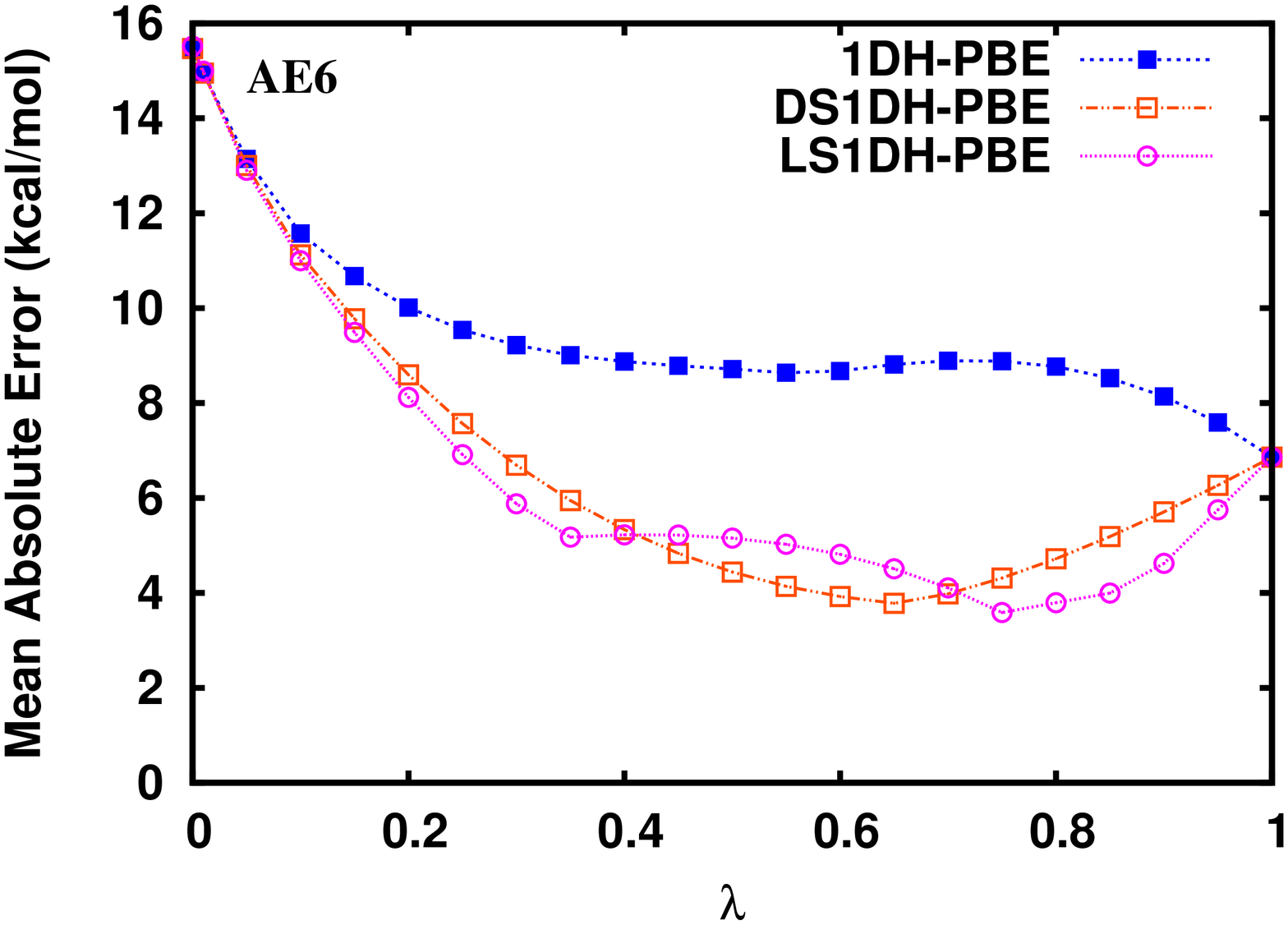}
\includegraphics[scale=0.3,angle=-90]{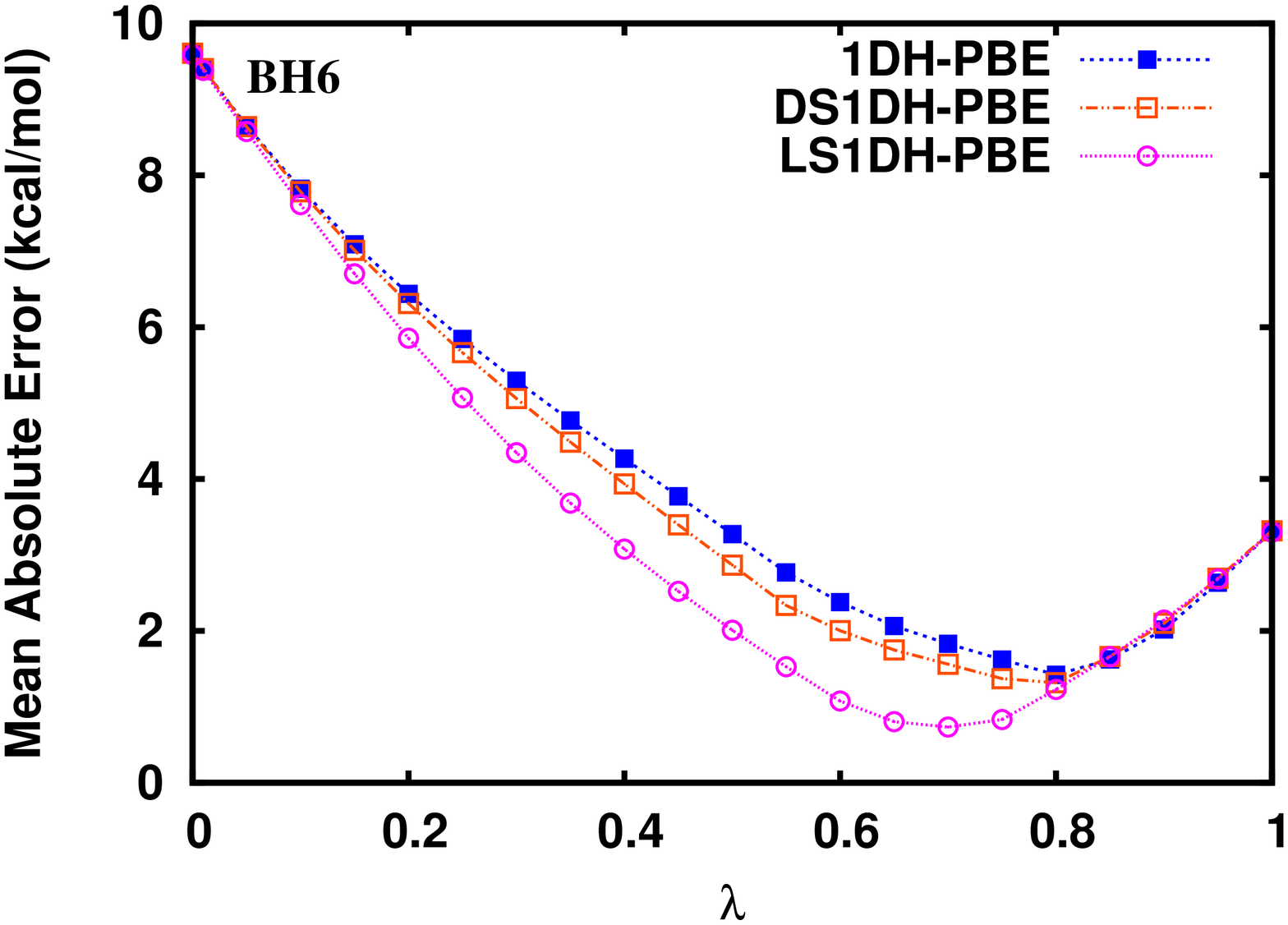}
\caption{(Color online) Mean absolute errors for the AE6 (left) and BH6 (right) test sets as functions of the parameter $\l$ for the 1DH, and DS1DH, and LS1DH approximations with the PBE exchange-correlation density functional. All calculations were carried out with the cc-pVQZ basis set.
}
\label{fig:AE6BH6}
\end{figure*}

Very recently, Br\'emond and Adamo~\cite{BreAda-JCP-11} have proposed a new class of double-hybrid approximations where the correlation functional is weighted by $(1-\l^3)$ and the MP2 correlation energy is weighted by $\l^3$, instead of $(1-\l^2)$ and $\l^2$, respectively. Applying this formula with the Perdew-Burke-Ernzerhof (PBE)~\cite{PerBurErn-PRL-96} exchange-correlation density functional, they have constructed the PBE0-DH double-hybrid approximation which performs reasonably well. In this work, we give a rationale for this class of double-hybrid approximations. For this, we start by recalling that the density-scaled correlation functional $E_c[n_{1/\l}]$ tends to the second-order G\"orling-Levy (GL2)~\cite{GorLev-PRB-93} correlation energy when the density is squeezed up to the high-density limit (or weak-interaction limit)
\begin{eqnarray}
\lim_{\l\to 0} E_c[n_{1/\l}] = E_c^{\text{GL2}},
\label{}
\end{eqnarray} 
which is finite for nondegenerate KS systems. The GL2 correlation energy can be decomposed as (see, e.g., Ref.~\onlinecite{Eng-INC-03})
\begin{eqnarray}
E_c^{\text{GL2}} = E_c^{\text{MP2}} + E_c^{\text{$\Delta$HF}},
\label{}
\end{eqnarray} 
where $E_c^{\text{MP2}}$ is the usual MP2 correlation energy expression
\begin{eqnarray}
E_c^{\text{MP2}} = -\frac{1}{4} \sum_{ij} \sum_{ab} \frac{\left| \bra{\phi_i \phi_j} \ket{\phi_a \phi_b}\right|^2}{\varepsilon_a + \varepsilon_b - \varepsilon_i - \varepsilon_j},
\label{EcMP2}
\end{eqnarray} 
with the antisymmetrized two-electron integrals $\bra{\phi_i \phi_j} \ket{\phi_a \phi_b}$, and $E_c^{\text{$\Delta$HF}}$ is an additional contribution involving the difference between the local multiplicative KS exchange potential $\hat{v}_{x}^{\KS}$ and the nonlocal nonmultiplicative HF exchange potential $\hat{v}_{x}^{\HF}$
\begin{eqnarray}
E_c^{\text{$\Delta$HF}} = - \sum_{i} \sum_{a} \frac{\left| \bra{\phi_i} \hat{v}_{x}^{\KS} - \hat{v}_{x}^{\HF} \ket{\phi_a}\right|^2}{\varepsilon_a - \varepsilon_i}.
\label{EcDHF}
\end{eqnarray} 
In both Eqs.~(\ref{EcMP2}) and~(\ref{EcDHF}), $\phi_k$ are the KS orbitals and $\varepsilon_k$ are their associated energies, and  the indices $i$,$j$ and $a$,$b$ stand for occupied and virtual orbitals, respectively. The single-excitation contribution $E_c^{\text{$\Delta$HF}}$ vanishes for two-electron systems, and in most other cases is negligible~\cite{Eng-INC-03}, so that the GL2 correlation energy is well approximated by just the MP2 contribution (evaluated with KS orbitals), $E_c^{\text{GL2}} \approx E_c^{\text{MP2}}$. This leads us to propose an approximation for $E_c[n_{1/\l}]$ based on a linear interpolation formula
\begin{eqnarray}
E_c[n_{1/\l}] \approx (1-\l) E_c^{\text{MP2}} + \l E_c[n].
\label{Ecscaledlinear}
\end{eqnarray} 
Plugging Eq.~(\ref{Ecscaledlinear}) into Eq.~(\ref{ExcDS1DH}), we directly arrive at what we call the {\it linearly scaled one-parameter double-hybrid} (LS1DH) approximation
\begin{eqnarray}
E^{\text{LS1DH},\l}_{xc} &=& \l E_x^{\HF} + (1-\l) E_x[n] 
\nonumber\\
&&+ (1-\l^3) E_c[n] + \l^3 E_c^{\text{MP2}},
\label{LS1DH}
\end{eqnarray} 
with the weights $(1-\l^3)$ and $\l^3$, thus giving a stronger rationale to the expression that Br\'emond and Adamo have proposed on the basis of different considerations. Further insight into this approximation can be gained by rewriting Eq.~(\ref{Ecscaledlinear}) in the alternative form
\begin{eqnarray}
E_c[n_{1/\l}] \approx E_c^{\text{MP2}} + \l \left( E_c[n] - E_c^{\text{MP2}} \right),
\label{Ecscaledexpand}
\end{eqnarray} 
which can then be interpreted as a first-order expansion in $\l$ around $\l=0$ with $E_c[n] - E_c^{\text{MP2}}$ approximating the third-order correlation energy correction $E_c^{(3)}[n]$ in G\"orling-Levy perturbation theory. In comparison, the zeroth-order approximation $E_c[n_{1/\l}] \approx E_c^{\text{MP2}}$ plugged in Eq.~(\ref{ExcDS1DH}) just gives the usual one-parameter hybrid (1H) approximation with the full correlation density functional~\cite{Bec-JCP-96,ErnPerBur-INC-96}
\begin{eqnarray}
E^{\text{1H},\l}_{xc} &=& \l E_x^{\HF} + (1-\l) E_x[n] + E_c[n].
\label{}
\end{eqnarray} 
In this sense, the LS1DH approximation of Eq.~(\ref{LS1DH}) can be considered as a next-order approximation in $\l$ beyond the usual hybrid approximation.

Figure~\ref{fig:hebe} illustrates the different approximations to the density-scaled correlation energy $E_c[n_{1/\l}]$ as a function of $\l$ for the He and Be atoms. The accurate reference curve is from the parametrization of Ref.~\onlinecite{ColSav-JCP-99}. For $\l=1$, it reduces to the exact correlation energy, while for $\l=0$ it is the GL2 correlation energy which tends to overestimate the correlation energy. In between these two limits, it is nearly linear with $\l$. The PBE correlation energy without density scaling [Eq.~(\ref{Ecnoscaled})] is the crudest approximation to $E_c[n_{1/\l}]$. The PBE correlation energy with density scaling (taken from the parametrization of Ref.~\onlinecite{ColSav-JCP-99}) gives a nearly linear curve. It is a fairly good approximation for He, but a less good approximation for Be where it underestimates the correlation energy for all $\l$, especially at $\l=0$. This is due to the presence of static correlation in this system that is not described by the PBE functional. Finally, the linear interpolation of Eq.~(\ref{Ecscaledlinear}) between MP2 (evaluated with PBE orbitals) and PBE appears as a good approximation for both He and Be. In fact, the linear interpolation is clearly the best approximation for Be, at least with the PBE functional.

For a more comprehensive comparison of the different approximations, we have performed calculations on the AE6 and BH6 test sets~\cite{LynTru-JPCA-03} with the 1DH, DS1DH, and LS1DH double hybrids for the PBE functional, using a development version of the MOLPRO 2010 program~\cite{Molproshort-PROG-10}. The AE6 set is a small representative benchmark set of six atomization energies consisting of SiH$_4$, S$_2$, SiO, C$_3$H$_4$ (propyne), C$_2$H$_2$O$_2$ (glyoxal), and C$_4$H$_8$ (cyclobutane). The BH6 set is a small representative benchmark set of forward and reverse barrier heights of three reactions, OH + CH$_4$ $\to$ CH$_3$ + H$_2$O, H + OH $\to$ O + H$_2$, and H + H$_2$S $\to$ HS + H$_2$. All the calculations are performed at the optimized QCISD/MG3 geometries~\cite{LynZhaTru-JJJ-XX}. We use the Dunning cc-pVQZ basis set~\cite{Dun-JCP-89,WooDun-JCP-93}. Core electrons are kept frozen in all our MP2 calculations. Spin-restricted calculations are performed for all the closed-shell systems, and spin-unrestricted calculations for all the open-shell systems. In Fig.~\ref{fig:AE6BH6}, we plot the mean absolute errors (MAE) for the two sets as a function of $\l$. For the AE6 set, the MAEs of the DS1DH and LS1DH approximations are quite similar, and are both much smaller that the MAE of the 1DH approximation for a wide range of $\l$. For the BH6 set, the LS1DH approximation gives a MAE which is significantly smaller than those of both DS1DH and 1DH for intermediate values of $\l$. 

As regards the choice of the parameter $\l$, the present data on the AH6 set gives an optimal value of $\l=0.75$ for the LS1DH double hybrid, with a minimal MAE of 3.59 kcal/mol, and for the BH6 set the optimal value is $\l=0.70$ giving a minimal MAE of 0.73 kcal/mol. However, the MAE is not very sensitive to the value of $\l$ around the optimal value, and Br\'emond and Adamo~\cite{BreAda-JCP-11} have argued for using $\l=0.5$ in defining the PBE0-DH approximation, using a similar argument as the one used by Becke for his ``half-and-half'' hybrid~\cite{Bec-JCP-93a}.

In summary, we have shown that the new class of double-hybrid approximations named here LS1DH [Eq.~(\ref{LS1DH})] can be understood as approximating the density-scaled correlation functional $E_c[n_{1/\l}]$ by a linear function of $\l$, interpolating between MP2 at $\l=0$ and a density-functional approximation at $\l=1$. Numerical results obtained with the PBE density functional confirms that the LS1DH approximation is a relevant double-hybrid model, and in fact tends to be more accurate than the DS1DH double-hybrid model [Eq.~(\ref{ExcDS1DH})] in which density scaling is applied to the PBE correlation functional. More generally, it can be expected that the LS1DH double-hybrid model will be more accurate than the DS1DH double-hybrid model when applied with a density-functional approximation that is inaccurate in the high-density limit. We hope that this work will help constructing other theoretically justified double-hybrid approximations.

\section*{Acknowledgments}
We thank A. Savin for discussions. This work was partly supported by Agence Nationale de la Recherche (ANR) via contract numbers 07-BLAN-0272 (Wademecom) and 10-BLANC-0425 (DinfDFT). Sanofi-Aventis is also acknowledged for financial support.


\end{document}